# International Journal Of Database Management System (IJDMS)

## Title Of Paper

"APPLICATION OF DATA MINING TECHNIQUES TO A SELECTED BUSINESS ORGANISATION WITH SPECIAL REFERENCE TO BUYING BEHAVIOUR"

## Authors :

1. Mrs. Tejaswini Abhijit Hilage.
   Qualification: MCA
   Assistant Professor,

   Deshbhakt Ratnappa Kumbhar College Of Commerce, Kolhapur.

   Address:

   148/A/4, 'Suverna', Aishwarya Park, Tanubai Nager, Behind Collector Office, Nagala Park, Kolhapur

   Cell: 7588576837, 0231-2663435

   Email: tejaswini.hilage@gmail.com

2. Dr. R.V. Kulkarni.

   Qualification : M.Sc. Ph.D.

   Professor & Head of department,

   Chh. Shahu Institute Of Business Education And Research, Kolhapur.

   Email : drrvkulkarni@siberindia.co.in

# Application of data mining techniques to a selected business organization with special reference to buying behavior

Mrs. Tejaswini Abhijit Hilage & Dr. R. V. Kulkarni.

Assistant Professor, Department Of Management, Deshbhakt Ratnappa Kumbhar College Of Commerce, Kolhapur - 416002

tejaswini.hilage@gmail.com

And

Professor & Head Of Department, Chh. Shahu Institute Of Business Education & Research Centre, Kolhapur – 416006

drrvkulkarni@siberindia.co.in

**ABSRTACT –**

*Data mining is a new concept & an exploration and analysis of large data sets, in order to discover meaningful patterns and rules. Many organizations are now using the data mining techniques to find out meaningful patterns from the database. The present paper studies how data mining techniques can be apply to the large database. These data mining techniques give certain behavioral pattern from the database. The results which come after analysis of the database are useful for organization. This paper examines the result after applying association rule mining technique, rule induction technique and Apriori algorithm. These techniques are applied to the database of shopping mall. Market basket analysis is performing by the above mentioned techniques and some important results are found such as buying behavior.*

**INTRODUCTION :**

Large databases maintained by retailers, telecom service providers & credit card companies contain very valuable information related to customers. These enterprises could benefit immensely in the areas of marketing, advertising and sales, if interesting and previously unknown customer buying and calling patterns can be discovered from the large volumes of data. Data mining helps marketing analysts to reflect actual behavior in different situations. Data mining is an attempt to source out pattern and trends in the data and infers rules from these patterns. With these rules the user will be able to support, review and examine decisions in some related business and scientific area. This opens up the possibility of a new way of interacting with databases.

An association rule is an expression of the form X=>Y, where X and Y are the sets of items. The goal is to discover all the rules that have the support & confidence greater than or equal to the minimum support and confidence respectively.

Let L = {l1, l2, l3… lm} be a set of literals called items.

Let D = database, be a set of transactions, where each transaction T is the set of items. T support an item x, if x is in T. T is said to support a subset of items x, if T supports each item x in X. Supports means how often X and Y occur together as a percentage of the total transactions. Confidence measures how much a particular item is dependent on another.

Apriori is a classical algorithm and is designed to operate on databases containing transactions. The theory of Apriori algorithm is that "All nonempty subsets of a frequent itemset must also be frequent."

Apriori principle is shown as below:

For all(x, y): (x belongs to y) => s(x) >= s(y).

Apriori Algorithm is as follow:

1. Let K=1.
2. Generate frequent itemsets of length 1.
3. Repeat until no frequent itemsets are identified.

Rule induction technique retrieves all interesting patterns from database.

In rule induction technique, the rule is of "if this then this". For example a rule that a supermarket might find in their data collected from scanners would be: "if pickles are purchased then ketchup is purchased'.

|  | Accuracy Low | Accuracy High |
|---|---|---|
| Coverage High | Rule is rarely correct but can be used often. | Rule is often correct and can be used often. |
| Coverage Low | Rule is rarely correct and can be only rarely used. | Rule is often correct but can be only rarely used. |

**REVIEW OF LITERATURE:**

Leonid Churilov, Adyl Bagirov, Daniel Schwartz, Kate Smith and Michael Dally [1] had already studied about combined use of self organizing maps & nonsmooth, nonconvex optimization techniques in order to produce a working case of a data driven risk classification system. Bagirov et al. [1]

propose the global optimization approach to clustering and demonstrate how the supervised data classification problem can be solved via clustering. Anthony D Anna & Oscar H. Gandy [2] develop a more comprehensive understanding of data mining by examining the application of this technology in the marketplace. As more firms shift more of their business activities to the web, increasingly more information about consumers and potential customers is being captured in web server logs. Anthony D Anna & Oscar H. Gandy [2] examine issues related to social policy that arise as the result of convergent developments in e_business technology and corporate marketing strategies.

Artificial neural networks are designed to model human brain functioning through the use of mathematics. Like neural network data mining through the use of decision tree algorithms discerns patterns in the data without being directed. According to Linoff "decision trees work like a game of 20 questions", by automatically segmenting data into groups based on the model generated when the algorithms were run on a sample of the data (1998, p. 44). Decision tree models are commonly used to segment customers into "statistically significant" groups that are used as a point of reference to make predictions (Vaneko and Russo, 1999). Both neural networks & decision trees require that one knows where to look in the data for patterns, as a sample of data is used as a training device. The use of market basket analysis & clustering techniques does not require any knowledge about relationships in the data, knowledge is discovered when these techniques are applied to the data. Market basket analysis tools sift through data to let retailers know what products are being purchased together. Clusters prove to be most useful when they are integrated into a marketing strategy.

Huda Akil, Maryann E. Martone, David C Van Essen [15] made a study about understanding the brain requires a broad range of approaches and methods from the domains of biology, psychology, chemistry, physics & mathematics. National Institute Of Health recently launched the Human Connectome Project and awarded grants to two consortia. The consortium led by Washington University of Minnesota aims to characterize whole brain circuitry & its variability across individuals in 1200 healthy adults. Neuroimaging & behavioral data from the HCP will be made freely available to the neuroscience community via a database & a platform for visualization & user friendly data mining.

Chandrika Kamath [14] in her study said that the size & the complexity of the data from scientific simulations, observations and experiments becoming a major impediment to their analysis. To enable scientists to address this problem of data overload and benefit from their improved data collecting abilities, the Sapphire project team has been involved in the research, development & application of scientific data mining techniques for nearly a decade. According to her team the raw data available for analysis was in the form of images, structured or unstructured mesh data with physical variables at each mesh point or time series data collected by different sensors. They designed and built a software toolkit with separate modules for different tasks such as denoising, background subtraction to identify moving objects in video, dimension reduction to identify key characteristics of objects, pattern recognition for clustering and classification.

Study made by Fadi Thabtah [13] about associative classification mining said associative classification integrates two known data mining tasks, association rule discovery and classification to build a model for the purpose of prediction. Another study is made by Tae Kyung Sung, Namsik Chang and Gunhee Lee about how data mining approach develop bankruptcy prediction model suitable for normal & crisis economic condition.The study made by HE Zengyou, XU Xiafei and DENG Shengchun [10] said that clustering is an important KDD technique with numerous applications, such as marketing & customer segmentation.

The aim of the study made by T. W. Rennie and W. Roberts [11] was to demonstrate the epidemiological use of multiple correspondence analyses as applied to tuberculosis (TB) data from North East Landon primary care trusts between the years 2002 & 2007 was used. TB notification data were entered for Multiple Corresponding Analysis allowing display of graphical data output. The study made by Shakil Ahmed, Frans Coenen and Paul Leng [4] consider strategies for partitioning the data to deal effectively. The study made by Balaji Padmanabhan & Alexander Tuzhilin [8] said previous work on the solution to analytical electronic customer relationship management problem has used either data mining or optimization methods, but has not combined the two approaches. By leveraging the strength of both approaches, the eCRM problems of customer analysis, customer interaction & the optimization of performance metrics can be better analyzed.

By considering previous studies authors find out the scope to go for research in market basket analysis using three different algorithms namely Association Rule Mining, Rule Induction Technique and Apriori Algorithm. Authors will make a comparative study of three techniques and adopt the best conclusion.

**OBJECTIVES OF STUDY :**

In view of foregoing discussion & considering the nature of present study, the authors has laid down following objectives :

- To develop software which will perform market basket analysis using data mining techniques.
- To suggest suitable data mining tool which will help the organization.
- To understand customer buying behavior with the help of data mining technique and tools.
- To study how data mining tech. work while mining the data from data warehouse.

**DATA COLLECTION METHODS :**

## Secondary Data :

1. Secondary data is collected from bills from the shopping malls.

2. Data is collected from periodic reports of shopping mall.

**DATA PRESENTATION AND DATA ANALYSIS :-**

Here authors has develop and analyzed the large database of shopping mall. For presentation purpose authors has shown sample bills from the database and applied different data mining techniques as below.

## Data Presentation :

Collected data is tabulated in transaction table :

**Transaction Table –**

**Table No. 1 =**

| Transaction ID | Items |
|---|---|
| 1 | Shabu, Parle G |
| 2 | Parle G, Sunflower oil |
| 3 | Shabu, Parle G, Sunflower oil, Pohe, Sugar, Rava, Dhoop |
| 4 | Sunflower oil, Suger, Rava, Parle G |
| 5 | Shabu, Maggi, Rava , Dhoop |
| 6 | Rava Suger, Agarbatti |
| 7 | Sunflower oil, Rava, Pen, Parle G |
| 8 | Suger, Rava, Pen |
| 9 | Shabu, Maggi, Agarbatti, Suger |
| 10 | Shabu, Suger, Rava, Sunflower oil. |
| 11 | Biscuit, Suger, Rava |
| 12 | Sunflower oil, Pen, Chocolate |
| 13 | Suger, Rava, Agarbatti |
| 14 | Sunflower oil, Suger, Rava, Notebook |
| 15 | Sunflower oil, Suger, Dhoop, Pohe. |

Select following items for calculation of association rules.

Sunflower, Suger, Rava, Dhoop.

**COADING :-**

Use binary 0 & 1.

0 indicate item is absent in the transaction &

1 indicate item is present in the transaction.

Input : Transaction Records :

**Table No. 2 =**

| Transaction ID | Sunflower | Suger | Rava | Dhoop |
|---|---|---|---|---|
| 1 | 0 | 0 | 0 | 0 |
| 2 | 1 | 0 | 0 | 0 |
| 3 | 1 | 1 | 1 | 1 |
| 4 | 1 | 1 | 1 | 0 |
| 5 | 0 | 0 | 1 | 1 |
| 6 | 0 | 1 | 1 | 0 |
| 7  | 1 | 0 | 1 | 0 |
| 8  | 0 | 1 | 1 | 0 |
| 9  | 0 | 1 | 0 | 0 |
| 10 | 1 | 1 | 1 | 0 |
| 11 | 0 | 1 | 1 | 0 |
| 12 | 1 | 0 | 0 | 0 |
| 13 | 0 | 1 | 1 | 0 |
| 14 | 1 | 1 | 1 | 0 |
| 15 | 1 | 1 | 0 | 1 |

**ANALYSIS –**

1. Analysis by Association Rule :

In association rule support and confidence is calculated. Some minimum support and confidence is assumed. The rules which will satisfies the criteria of minimum support and minimum confidence will be true otherwise false.

Support and confidence are calculated as below :-

Support (x -> y) = (Number of transactions containing x & y) / (Total no. of transactions.)

Confidence (x -> y) = (Support of x -> y) / (Support of x)

Assume minimum support = 50% and minimum confidence = 70 %

**Table No. 3 =**

| Sr. Number | X | Y | Support (%) | Confidence (%) | Is in rule? |
|---|---|---|---|---|---|
| 1 | Sunflower oil | Suger | 31 | 71 | False |
| 2 | Sunflower oil | Rava  | 31 | 71 | False |
| 3 | Sunflower oil | Dhoop | 12 | 29 | False |
| 4 | Suger | Rava  | 56 | 75 | True  |
| 5 | Suger | Dhoop | 12 | 17 | False |
| 6 | Rava  | Dhoop | 12 | 18 | False |

Following association rules are found:-

1. If sunflower oil then Suger

    For this rule support = 31 % & confidence = 71 %.

    The rule does not satisfies the criteria of minimum support, this rule is false.

    So we can say that when customer buys sunflower, he will not buy sugar.

2. If sunflower oil then Rava

    Here,

    Support = 31 %, confidence = 71.

    As it does not satisfies the criteria of min. support, the rule is false.

    So when customer buys sunflower oil, he will not buy rava.

3. If sunflower oil then Dhoop.

    Here,

    Support = 12 % and confidence = 29 %.

    The rule does not satisfies the criteria of minimum support and minimum confidence, so this rule is false.

    So when customer buys sunflower oil, he will not go for Dhoop.

4. If Sugar then Rava.

    Here,

    Support = 56 % and confidence = 75 %.

    Here, the rule satisfies both the criteria of minimum support and minimum confidence.

    So, the rule is true.

    Therefore when customer buys Sugar, he will buy Rava with 75% confidence.

5. If Sugar then Dhoop.

    Here, support = 12 % & confidence = 17 %.

    The rule does not satisfy the criteria of minimum support & min. confidence.

    So, this rule is false.

    Hence we can predict that when customer buys sugar, he will not buy Dhoop.

6. If Rava then Dhoop.

    Here, support = 12 % and confidence = 18 %.

    This rule does not satisfy the criteria of min. support & min. confidence.

    So, the rule is false.

    So we can predict that when customer buys Rava, he will not go for Dhoop.

## ANALYSIS BY RULE INDUCTION TECHNIQUE :

In rule induction technique accuracy & coverage is calculated. Some minimum accuracy and coverage is assumed. The rules satisfying minimum accuracy and coverage is true otherwise false.

Accuracy and coverage is calculated as below :

Accuracy (x -> y) = (Number of transaction containing x and y) / (Number of transaction with x)

Coverage (x -> y) = (Number of transaction containing x) / (Total number of transactions)

Assume minimum accuracy = 50 % and minimum coverage = 70 %.

**Table No. 4 =**

| Sr. Number | X | Y | Accuracy (%) | Coverage (%) | Is in rule? |
|---|---|---|---|---|---|
| 1 | Sunflower oil | Suger | 71 | 44 | False |
| 2 | Sunflower oil | Rava | 71 | 44 | False |
| 3 | Sunflower oil | Dhoop | 29 | 44 | False |
| 4 | Suger | Rava | 75 | 75 | True |
| 5 | Suger | Dhoop | 17 | 75 | False |
| 6 | Rava | Dhoop | 18 | 69 | False |

Following association rules are found:

1. If sunflower oil then Suger

    For this rule accuracy = 71 % & coverage = 44 %.

    As this rule satisfies the criteria of minimum accuracy but does not satisfies the criteria of minimum coverage, so this rule is false.

    So when customer buys sunflower, he didn't go for sugar.

2. If sunflower oil then Rava

    Here,

    Accuracy = 71 %, coverage = 44%.

    As it satisfies the criteria of min. accuracy but does not satisfies the criteria of minimum coverage, this rule is false.

    So when customer buys sunflower oil, he didn't go for rava.

3. If sunflower oil then Dhoop.

    Here,

    Accuracy = 29 % and coverage = 44 %.

    The rule does not satisfies the criteria of minimum accuracy and minimum coverage.

    So this rule is false.

So when customer buys, he didn't go for Dhoop.

4. If Sugar then Rava.

    Here,

    Accuracy = 75 % and coverage = 75 %.

    Here, the rule satisfies the criteria of minimum accuracy and minimum coverage.

    So, the rule is true.

    Therefore when customer buys Sugar, he will buy Rava with 75% confidence.

5. If Sugar then Dhoop.

    Here, accuracy = 17 % & coverage = 75 %.

    The rule does not satisfy the criteria of minimum accuracy, so this rule is false.

    Hence we can predict that when customer buys sugar, he didn't go for Dhoop.

6. If Rava then Dhoop.

    Here, accuracy = 18 % and confidence = 69 %.

    The rule does not satisfy the criteria of minimum accuracy & minimum coverage.

    So, the rule is falses.

    So we can predict that when customer buys Rava, he will not go for Dhoop.

## ANALYSIS BY APRIORI ALGORITHM :-

The theory of Apriori algorithm is that prune the items with lowest support count until you get the one record with largest support count.

Support count is calculated as below :

Support count (x) = Number of transactions containing x.

In Apriori algorithm two tables C and L are prepared. Support count is calculated for each item and recorded in C.

In L the items having lowest support count is deleted.

Prepare table C1 and L1

First C1:

**Table No. 5 =**

| Item | Support count |
|---|---|
| Sunflower oil | 7 |
| Sugar | 12 |
| Rava | 11 |
| Dhoop | 3 |

In C1,

Support count for Sunflower Oil = 7, that indicates there are 7 bills containing sunflower oil.

Support count for Suger = 12, means there are 12 bills containing Suger.

Support count for Rava = 11, means there are 11 bills containing Rava.

Support count for Dhoop = 3, indicate there are 3 bills which contains Dhoop.

L1:

**Table No. 6 =**

| Item | Support count |
|---|---|
| Sunflower oil | 7 |
| Suger | 12 |
| Rava | 11 |

In L1, the items having lowest support count in C1, are deleted. (ie. Pruning process)

Here,

Dhoop are having lowest support count = 3.

In L1 Dhoop is deleted.

Hence,

L1 contains Sunflower Oil, Suger & Rava with 7,12 and 11 support count respectively.

Now Prepare C2 :

C2:

**Table No. 7 =**

| Item | Support count |
|---|---|
| Sunflower oil, Suger | 5 |
| Sunflower oil, Rava | 5 |
| Sunflower oil, Dhoop | 2 |
| Suger, Rava | 9 |
| Suger, Dhoop | 2 |
| Rava, Dhoop | 2 |

In C2, algorithm will calculate support count for between two items.

Here,

Support count for Sunflower oil and Suger = 5, indicates there are 5 bills which contains Sunflower oil and Suger.

Support count for Sunflower oil & Rava = 5, indicates there are 5 bills which contains Sunflower oil and Rava.

Support count for Sunflower oil & Dhoop = 2, means there are two bills which contains Sunflower oil & Dhoop.

Support count for Suger & Rava = 9, means there are 9 wills which contains Suger & Rava.

Support count for Suger & Dhoop =2, indicates there are 2 bills which contains Suger and Dhoop.

Support count for Rava and Dhoop = 2, indicates there are 2 bills which contains Rava and Dhoop.

Now, L2 :

**Table No. 8 =**

| Item | Support count |
|---|---|
| Sunflower oil, Suger | 5 |
| Sunflower oil, Rava | 5 |
| Suger, Rava | 9 |

In L2, prune the items having lowest support count. Here items having support count = 2 are deleted from C2.

Now,

Calculate C3 :

**Table No. 9 =**

| Item | Support Count |
|---|---|
| Sunflower oil, Suger, Rava | 4 |
| Sunflower oil, Suger, Dhoop | 2 |
| Sunflower oil, Rava, Dhoop | 1 |
| Suger, Rava, Dhoop | 1 |

In C3, calculate support count between three items.

In C3,

Support count for Sunflower oil, Suger & Rava = 4, indicates there are 4 bills which contains Sunflower Oil, Suger and Rava.

Support count for Sunflower oil, Suger and Dhoop=2, indicates there are 2 bills which contains Sunflower Oil, Suger & Dhoop.

Support count for Sunflower oil, Rava and Dhoop =1, indicates there is one bill which contains Sunflower oil, Rava and Dhoop.

Support count for Suger, Rava & Dhoop = 1, indicates there is one bill which contains Suger, Rava and Dhoop.

Now, L3 =

**Table No. 10 =**

| Item | Support count |
|---|---|
| Sunflower oil, Suger, Rava | 4 |

In L3 we have deleted the items having lowest support count in C3.

So, L3 will contain Sunflower Oil, Suger and Rava with support count 4 and with Support = 26% and confidence = 78%

So from Apriori Algorithm we can predict that, the customer who will buy Sunflower Oil and Suger can go for Rava with 78 % confidence.

By observing calculation from the Association Rule, Rule Induction Technique and Apriori Algorithm, following result is found :

Sunflower Oil, Suger & Rava are strongly associated with each other.

Authors have shown the analysis for four products but the developed software is able to analyze n number of products.

**FINDINGS AND SUGGESTIONS :**

**FINDINGS :**

The data of various transactions made by customers is stored in the data base. This database gives certain meaningful information as below :

1. The results show that for this organization Sunflower Oil, Suger and Rava are strongly associated with each other.

2. There are some buying habits of customers.

3. There is some association exist between the different products which are purchased by the customer in single trip. And it is observed that customer buys those products which are highly associated with each other.

**SUGGESTIONS :**

On the basis of the results from data mining tool the following suggestions are given :

1. Organization can place Sunflower Oil, Suger and Rava near to each other for increasing sell of these products.
2. As organization have data of all transactions made by customer, so organization can install such data mining tools which will generate association rules for different products.
3. By considering the association rules generated by data mining tool, organization can rearrange their shelf for promotion of sale.

**REFERENCES :-**